\documentclass[12pt]{article} 
\usepackage{amsfonts}
\usepackage{amssymb}
\usepackage{amsmath}
\usepackage{amsthm}
\usepackage{setspace}
\usepackage{color}
\usepackage{cite}
\usepackage[dvipsnames]{xcolor}
\definecolor{cobalt}{rgb}{0.0, 0.28, 0.67}
\definecolor{darkblue}{rgb}{0.0, 0.0, 0.55}
\definecolor{darkpowderblue}{rgb}{0.0, 0.2, 0.6}
\definecolor{egyptianblue}{rgb}{0.06, 0.2, 0.65}
\usepackage{setspace}
\usepackage{xcolor,colortbl}
\usepackage[pdfencoding=auto]{hyperref}
\hypersetup{
    colorlinks=true,
    linkcolor=egyptianblue,
    citecolor=egyptianblue,
    filecolor=magenta,      
    urlcolor=darkblue,
}
\hypersetup{pdfstartview={XYZ null null 1}}
\usepackage[]{natbib}
\usepackage{pict2e}
\usepackage{tikz}
\graphicspath{{graphs/}}
\usepackage{pgfplots}
\usepackage[font=footnotesize, labelfont=bf]{caption}
 \usepackage{booktabs}
\usepackage{pgflibraryarrows}
\usepackage{pgflibrarysnakes}
\usepackage[font=scriptsize]{subcaption}
\usepackage{threeparttable}
\usepackage{dcolumn}
\newcolumntype{v}{D{.}{.}{2.4}}
\newcolumntype{f}{D{.}{.}{2.2}}
\newcolumntype{p}{D{.}{.}{-1}}
\usepackage{graphicx}
\usepackage{epstopdf}
\usepackage{parskip}
\usepackage{geometry}
\usepackage{lscape}
\usepackage{array}
\usepackage{multirow}
\usepackage{indentfirst}
\usepackage{paralist}
\usepackage{sectsty}
\usepackage{float}
\usepackage{placeins}
\usepackage{threeparttable}
\usepackage{url}
\sectionfont{\centering}
\usepackage[compact]{titlesec}
\theoremstyle{plain}

\theoremstyle{definition}

\titleformat*{\section}{\scshape\large\color{darkblue}}
\titleformat*{\subsection}{\scshape\large\color{darkblue}}

\begin{document}

\title{\Large 
 The Forest Behind the Tree:\\ 
 Heterogeneity in How US Governor's Party Affects Black Workers\thanks{The authors thank Louis-Philippe Beland for making the data publicly available and Thomas (Danny) Boston for comments on the early version of this manuscript. All errors and omissions are ours.}}
\author{\normalsize Guy Tchuente\thanks{School of Economics, University of Kent, NIESR, and GLO, UK; Correspondence Email: g.tchuente@kent.ac.uk.}
\and \normalsize Johnson Kakeu\thanks{ Department of Economics, University of Prince Edward Island, Canada, Email:
j.kakeu@upei.ca.}     
 \and
\normalsize John Nana Francois\thanks{College of Business, West Texas A\&M University, Canyon, USA, Email: jfrancois@wtamu.edu.}  }

\maketitle
\doublespacing
\thispagestyle{empty}

\begin{abstract} \onehalfspacing \noindent 
Income inequality is a distributional phenomenon. This paper examines the impact of U.S governor's party allegiance (Republican vs Democrat)  on ethnic wage gap.  A descriptive analysis of the distribution of yearly earnings of {\em Whites} and {\em Blacks} reveals a divergence in their respective shapes over time suggesting that aggregate analysis may mask important heterogeneous effects. This motivates a granular estimation of the comparative causal effect of governors' party affiliation on labor market outcomes.  We use a regression discontinuity design (RDD) based on marginal electoral victories and samples of quantiles groups by wage and hours worked.  Overall, the distributional causal estimations show that the vast majority of subgroups of black workers earnings are not affected by democrat governors’ policies, suggesting the possible existence of structural factors in the labor markets that contribute to create and keep a wage trap and/or hour worked trap for most of the subgroups of black workers. Democrat governors increase the number of hours worked of black workers at the highest quartiles of earnings. A bivariate quantiles groups analysis shows that democrats decrease the total hours worked for black workers who have the largest number of hours worked and earn the least. Black workers earning more and working fewer hours than half of the sample see  their number of hours worked increase under a democrat governor.\vspace{.5cm}

\vspace{.5cm}

\vspace{.5cm}

\end{abstract}
\newpage
\setcounter{page}{1}
\onehalfspacing
\maketitle
\doublespacing
\section{Introduction}
 
Racial income inequality is a distributional phenomenon that is drawing a growing concern on both sides of the political aisle in the U.S.  In this paper, we address one question related to the extent to which party allegiance affects the labor market outcomes for blacks relative to other demographic groups in the U.S. More specifically, we investigate the causal impact of the gubernatorial party allegiance (i.e.,  Democratic or Republican) of US governors by scrutinizing its effects on subgroups of the population of  Black  workers in the United States. The effect of the governors' party is assessed by decomposing the sample into different quantiles based on the level of earnings and number of hours worked--- two  important labor market outcomes that matter for wealth accumulation.  The motivation for this decomposition are twofold: First, a descriptive analysis of the earning distribution of whites and blacks in the U.S uncovers that the earning distribution of whites is bi-modal while that of blacks stays uni-modal (see Figure~\ref{fig: race_only} on page~\pageref{fig: race_only}). Given the difference in the shape of the two distributions, estimating the governor's party affiliation effect for the average representative black American may lead to effects that are substantially different from the ones  richer, or more disadvantaged individuals are experiencing. This difference is consistent with the existence of a subtended increased in wage inequality in the 80's and 90's described by \cite{autor2008trends}.  While the first motivation is driven by statistical concerns, the possibility of facing a highly heterogeneous causal effect when evaluating the policies is highlighted by several papers (see \cite{autor2008trends}, \citep{bake18}). The second motivation of this study stems from the growing literature showing that some policies have a different impact for individuals having higher income and those having lower income \citep{bake18}.

\begin{table}[htpb!]
\centering  
\caption{Causal effect of governor Party Affiliation on Number of Hours worked by Blacks}\label{tab:VisH}
\begin{tabular}{|c|c|c|c|c|c|}
\hline
   & $Q_1$ Earnings                    &  $Q_2$ Earnings &  $Q_3$ Earnings                    &  $Q_4$ Earnings                   &  Total Hours \\ [0.4em] \hline
$Q_1$ Hours &                        &    & \cellcolor{green!50}\textbf{+} &                        & \cellcolor{green!50}\textbf{+} \\ [0.6em] \hline 
$Q_2$ Hours &                        &    &                        & \cellcolor{green!50}\textbf{+} &  \\ [0.6em] \hline
$Q_3$ Hours & \cellcolor{red!60}\textbf{-} &    &                        &                        &  \\ [0.6em] \hline
Total Earnings   &                        &    &                        &                     \cellcolor{green!50}\textbf{+}   & \cellcolor{green!50}\textbf{+} \\ [0.6em] \hline
\end{tabular}
\begin{tablenotes} \footnotesize
\item {\em Notes:} This table summarizes the qualitative results of the estimation of governor's party effects using RDD in different quartiles of earnings and hours worked. The cells left blank are subgroups with non statically significant effects at levels 1, 5, or 10\%. The cells in green are subgroups with positive and statistically significant effects, while red cells are negative effects. 
\end{tablenotes}
\end{table}

This paper brings new insights into the existing literature by showing that governor party allegiance plays different role for different subgroups of the population. Our analysis uses labor market outcome, the total yearly earnings and  number of hours worked, and creates quantiles subgroups along them. 

\begin{table}[H]
\centering 
\caption{Causal effect of governor Party Affiliation on Blacks' Earning} \label{tab:VisE}
\begin{threeparttable}

\begin{tabular}{|l|l|l|l|l|c|}
\hline
   & $Q_1$ Earnings                     & $Q_2$ Earnings  & $Q_3$ Earnings                     & $Q_4$ Earnings                      & Total Hours \\ \hline
$Q_1$ Hours &                        &    &      &                        &  \\ \hline
$Q_2$ Hours &                        &    &                        &  &  \\ \hline
$Q_3$ Hours &   &    &                        &                      &  \\ \hline
 Total Earnings    &                        &    &                        &                        &  \cellcolor{green!50}\textbf{+}\\ \hline
\end{tabular}
\begin{tablenotes} \footnotesize
\item {\em Notes:} This table summarizes the qualitative results of the estimation of governor's party effects using RDD in different quartiles of earnings and hours worked. The cells left blank represent subgroups with a non statically significant effects at levels 1, 5, or 10\%. The cells in blue are subgroups with positive and statistically significant effects, while red cells are negative effects.  It should be noted that in none of the subgroups the party affiliation of the governor has any effect on the outcome except for the full sample.
\end{tablenotes}

\end{threeparttable}
\end{table}

We first document the growing, over-time, divergence in the distribution of annual wage of blacks and whites workers. This difference in the distribution is then tested using the non-parametric Kolmogorov-Smirnov test.  The evidence differences in the distribution motivate our estimation method that combines a Regression Discontinuity Design (RDD) and decomposition of the sample by quantiles approach. We exploit marginal victories at the state gubernatorial elections to design an RDD to estimate the effect of the governor on black's labor market outcomes and use subgroups granular analysis to unveil a large heterogeneity in the treatment effects.

As summarized in tables \ref{tab:VisH} and \ref{tab:VisE}, the granular analysis of the causal effect of governors suggests that there is no difference between Republican and Democrat governors in most quantiles groups, especially for earning. As for hours worked, Democrat governors increase the participation of black workers at the highest quantiles of earning. Democrats governors decrease the total hours worked for black workers who have the largest number of hours and earn the least while increasing the number of hours of black workers earning more and working fewer hours than half of the sample.  We also find that in the US, independent of the party affiliation of the governor,  certain subgroups of black workers  see no change in their labor market condition, suggesting the existence of a low-wage trap for lower-income quantiles individuals. The low-wage trap may be viewed as a situation where those black people already in low-paid work are not able to increase income no matter which party is in charge of making gubernatorial policies. One work is closely related to \cite{beland2015political} who showed that party allegiance matters with Democratic governors causing an increase in the annual hours worked by blacks relative to whites. While insightful, the methodology used in that work is silent on the income and worked hours distribution heterogeneity in investigating the question of the causal effect of party allegiance.  The analysis is based on the first moments of the whole distribution of the black population, namely the average value.  Several studies have shown that the analysis based on the average value or the median of the whole distribution may be misleading when the shape of the distribution is not well-behaved \citep{ banz19}. For instance, in surveying studies that look at race differences in analyzing the effect of pollution, \cite{ banz19} conclude that the relationship estimated from aggregated data is only equal to the relationship at the micro-level if there are no group-level effects correlated with the variable of interest.  Therefore, it is increasingly suggested to go beyond the aggregate average indicator analysis to a heterogeneous approach that explores finer sub-groups when analyzing race inequality and labor market issues. For instance, \cite{man18} showed that the observed stable ratio between median black and median white family incomes conceals two large and diametrically opposed trends between 1968 to 2016.  Along the same lines, the work by \cite{bake18} discusses why the median black man’s relative position in the earnings distribution has remained constant in the overall distribution over the past 70 years. They go beyond the standard mean or median to use the concept of rank earning gap that provides a finer picture of heterogeneity issues in several subgroups of the income distribution.

As mentioned earlier, empirically assessing the extent to which subgroups' considerations influence causal inference can increase our understanding of the link between global average effects and more micro effects.  The average effect may not be a meaningful statistic if the application of the analysis to a subgroup in the analysis of the effects does not lead to the same conclusion as the average effect on the whole distribution \citep{ber20}. One important motivation to scrutinize in going beyond macro aggregate estimation to less aggregate analysis is the increasing public demand for more robust approaches to the evaluation of public policy programs directed toward specific subgroups of the population \citep{he01,wi21}. There is a small but growing strand of literature that is looking at these issues in using Regression discontinuity design approaches. For instance, \cite{fra12} use quantile treatment effects in the regression discontinuity designs design for studying the distributional effects of a children’s education program. Their data exhibits regular distributions, excluding the possibility of dealing with isolated masses in the analysis, unlike our income and hours worked data. Along the same lines, \cite{huan20} provide a methodology for performing causal inferences that combine both a local quantile regression analysis and a regression discontinuity design.

The remainder of this paper unfolds as follows: In Section 2, describes some background by explaining the crucial role of the governor in shaping the US labor market.  Section 3 presents some descriptive statistics of the data, while section 4 discusses the results obtained from the causal inference. 

\section{The role of US governors in the shaping of states’ labor markets}

The US political system grants governors with a high degree of power over the management of the rules that shape the states’ economies, including labor market conditions. The governor heads the executive branch in each state, sets policies, prepares and administers a budget, recommends legislation, signs laws, and appoints department heads. Governors can veto state bills. Unlike U.S. presidents, many governors also have additional veto powers at their disposal. For instance, most states provide governors the power of the line-item veto, which gives governors the ability to strike out a line or individual portions of a bill while letting the remainder pass into law.  However, party composition at the state legislature has consequences on how governors will implement their own policy agendas. Alternatively, when the governor is not from the same party as the one controlling the state, she/he may have to work harder to build relationships and to reach a consensus.  In the situation known as unified government where the governor’s party affiliation controls the legislature, governors are hamstrung in their ability to implement their policy choices.   Under the leadership of the governors, states can set taxes, manage budgets, and regulate businesses. They can also enforce worker protections by regulating issues related to minimum wage, overtime pay coverage, anti-discrimination protections, workers' compensation, unemployment insurance, paid sick leave, and paid family leave. Other state activities that are important in shaping the labor market include managing the healthcare system, funding the public sector retirement plans, increasing access to affordable home care, and providing childcare services.  The way these policies are implemented is of particular importance for people working in low-paid jobs including janitorial, delivery, home care, agriculture, landscaping, security, hospitality, trucking, transportation, and warehousing. The substantial role governors play suggests that their gubernatorial party affiliation could have a pivotal role in shaping the labor market allocations.  

\section{Data and Descriptive Analysis} 

The data used in this paper are obtained by matching gubernatorial elections and the data from the Current Population Survey’s (CPS’s) March supplements from 1977 to 2008 as constructed by \cite{beland2015political}.\footnote{For detail on the construction of data set see \cite{beland2015political} in Section III.}

\subsection{Comparing Income Distribution by Ethnicity}
The large majority of work on the racial difference in labor outcomes has focused on mean or median differences (\cite{bake18}, \cite{che20}). By focusing on measures of central tendency, there is a possibility that the measure of the effect does not fully capture the reality in the context of US racial wealth inequality. In a recent paper, \cite{bake18} have identified some economics reasons to be interested in the full distribution of income when analyzing the racial wage gap: the disproportionate recent increase in non-work among U.S black men,\footnote{The over representation of black men out of labor market exacerbated the need to account for all men  to  avoid sample selection issues.} and recent evidence suggesting that economic changes such as raising general inequality differently affected black outcomes at different points in the distribution. For similar reasons, our analysis will take a holistic approach by examining the causal effect at all parts of the distribution.

\begin{figure}[hbt!]
    \centering
        \caption{Distribution of Income by Race }
    {\includegraphics[scale=1.4]{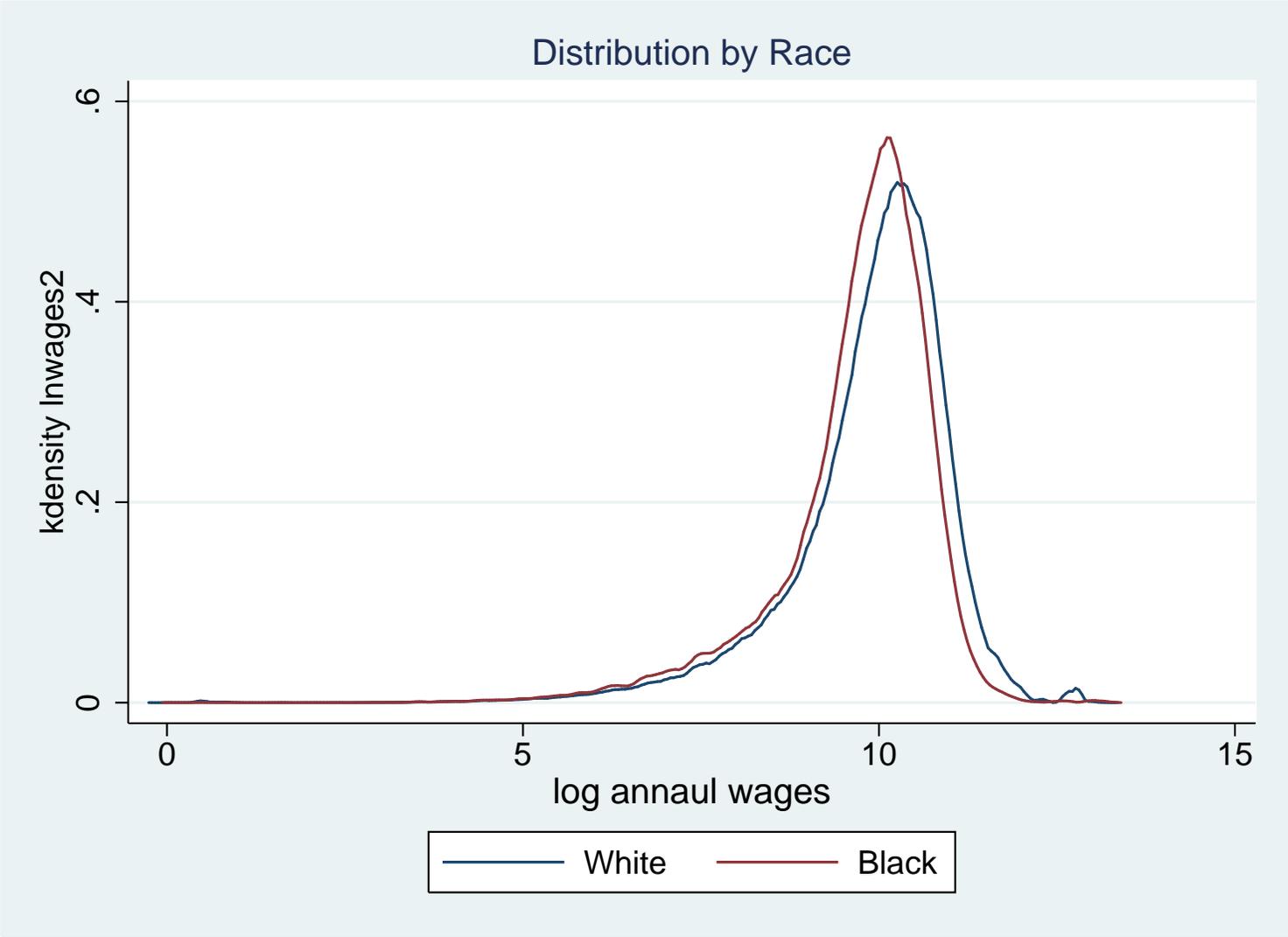}}
    \label{fig: race_only}
\end{figure}

\newpage

\begin{figure}[H]
\begin{center}
   {\caption{Distribution of Income by Race 4-year split 1977--1992 }
    \resizebox{11cm}{6cm}{\includegraphics[scale=1]{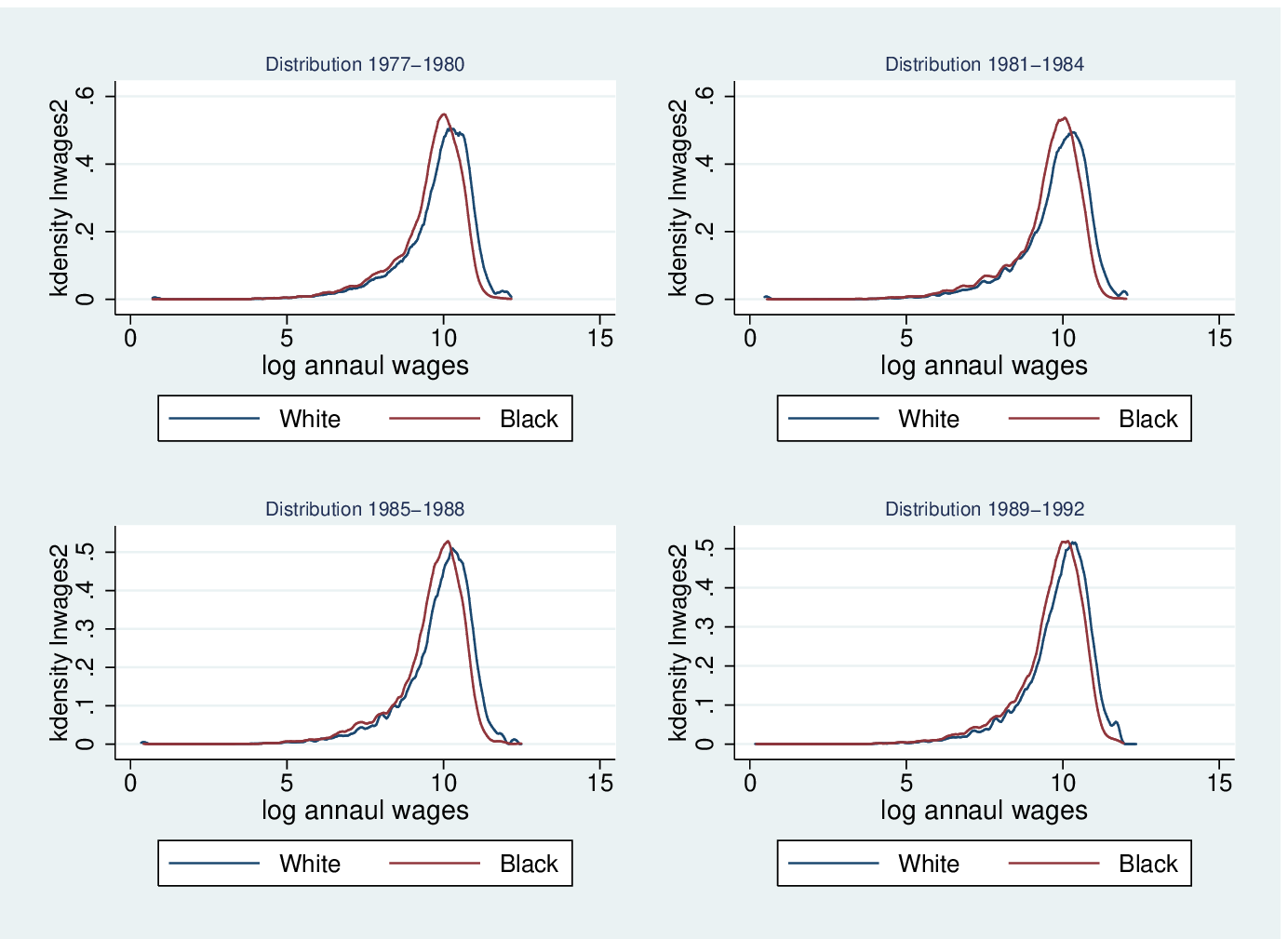}}
     \label{fig: race2_92}}\\
\end{center}

The analysis of figures \ref{fig: race_only} and \ref{fig: race2_92}  as well as figures \ref{fig: race_onlyd}  to \ref{fig: race1_08} [in Appendix~\ref{Hallelujah}] produce statistical  evidence supporting fundamental difference (beyond mean or median) in the income distribution by race. Each figure represents the distribution of the natural logarithm of individual annual income by ethnicity.\\

\begin{center}{
\caption{Distribution of Income by Race 4-year split 1993--2008}
    \resizebox{11cm}{5.7cm}{\includegraphics[scale=1]{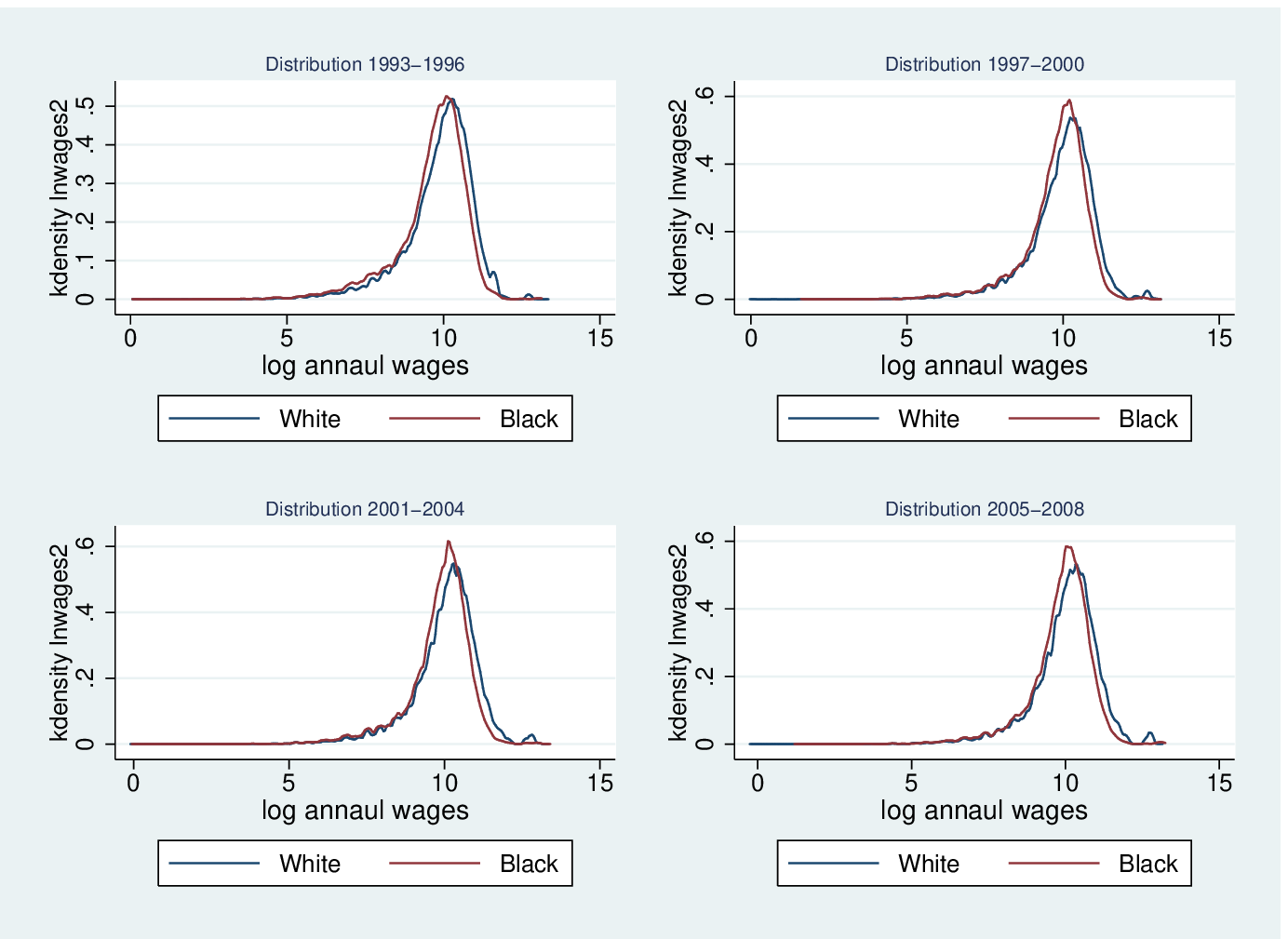}}
     \label{fig: race2_08}}
     \end{center}
\end{figure}

 Figure \ref{fig: race_only} shows that the income distribution of whites is bi-modal while  of blacks is uni-modal. A Kolmogorov-Smirnov test for  difference in the distribution confirms that at both distributions are different with the white individuals distribution more likely to have larger values.\footnote{The results of the others Kolmogorov-Smirnov tests are similar.}

Figures \ref{fig: race2_92} and \ref{fig: race2_08} present the dynamic of the distributions over time. While from 1977 to 1992 the distribution of  whites income dominates that of blacks, it is only from 1993 that the shapes of the distribution start to diverge. As we move from 1993 to 2008, the whites' income distribution becomes more bi-modal. This means the level of income inequality become more pronounce for whites than for blacks.\footnote{\cite{ heyw12} provide empirical evidence that the racial wage gap is the largest at the very top of the distribution of pay components related to bonuses, commissions, and so forth. The authors, however, do not address the question as to whether bonuses and commissions can be used to discriminate against an employee because of their sex, race, age, disability, or religion. While there is suggestive evidence that non-black managers hire more whites and fewer blacks than do black managers \citep{giu09}, there is also no obligation on employers to make the criteria for awarding bonuses transparent.}  Figures \ref{fig: race2_08} to \ref{fig: race_onlyd} [ in Appendix~\ref{Hallelujah} ]suggest that the observed divergence in income distribution by ethnicity is present irrespective of the party allegiance of the ruling state's Governor. 

To the extent that the distribution of income exhibits substantial differences and they are existing evidence of differential effects of the policy depending on the position of a black worker in the distribution, we propose a holistic approach to the causal analysis of political parties in labor market outcome in the US. 

\subsection{Evidence of Discontinuity by Quantiles Subgroups}

The application of the RDD methodology for each quantile implies that we need to investigate the presence of discontinuity at 0 percent at each income quantiles subgroup.  The Figure  \ref{fig:rdgraph_hours}, and figures   \ref{fig:rdgraph_worked}, and \ref{fig:rdgraph_wage} [in Appendix~\ref{Hallelujah}]  have four panels, each of them representing a quantile of annual income.  Each panel shows the discontinuity at 0 percent, this represents a Democratic governor barely marginal win over a Republican. Figure \ref{fig:rdgraph_hours} represents the hours worked by white and black workers. Figure  \ref{fig:rdgraph_worked} and \ref{fig:rdgraph_wage}  in Appendix~\ref{Hallelujah} represent respectively the proportion of employed and the natural logarithm of earning of whites and blacks.  

\begin{figure}[htpb!]
\caption{Margin of Democratic Victory and Total Hours Worked per Year for black (left on each panel) and white (right on each panel). The analysis is by Earning quartile.}
    \centering
    \includegraphics{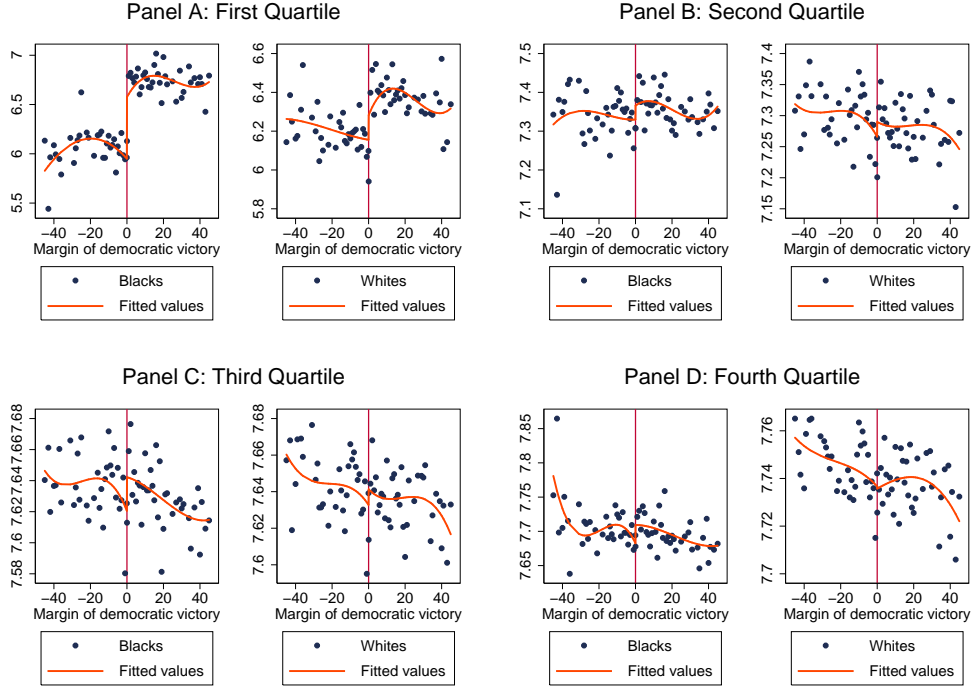}
    \label{fig:rdgraph_hours}
\end{figure}

Figure \ref{fig:rdgraph_hours}  suggests a positive effect of having a Democrat governor on the number of hours worked by workers. The effect seems to be concentrated at the lowest quartile. While the effect looks stronger for blacks it also exists for white in the first quartile. The heterogeneity in the evidence of the presence of discontinuity at the cut-off supports the need for an in-depth analysis of the effect of political parties.

The other labor market outcomes of interest in this section's analysis are the probability of employment represented in Figure \ref{fig:rdgraph_worked} and the the annual total earning shown in Figure \ref{fig:rdgraph_wage}. The analysis of figures \ref{fig:rdgraph_worked} and \ref{fig:rdgraph_wage} suggests that a Democrat victory has a positive effect on the probability of work and earning when individuals in the first and second quartiles are considered. The effects are larger for black workers and virtually no effect for white workers.  Moreover, as we move to higher quartiles, electing a Democrat governor does not affect the extensive margin of working irrespective of the race.

The results suggest a certain level of heterogeneity in the effect of governors' party allegiance on labor market outcomes. Individuals with earnings in the lower quartiles (first and second) seem the most affected by the marginal election of a democrat vs a republican governor. It is worth noting the fitted values obtained in Figures \ref{fig:rdgraph_hours} and  figures  \ref{fig:rdgraph_worked} and \ref{fig:rdgraph_wage} [in Appendix~\ref{Hallelujah}] are estimated without controlling for any covariates.  The next section extends this analysis by introducing additional controls for better precision and by dividing the sample into quartiles along two dimensions (annual income and hours worked). This also enable us to understand more precisely the extend of the heterogeneity of the governors effect on labor market outcome.

\section{Methodology for The Causal Estimation}

The effects of interest are identified based on the application of the regression discontinuity design (RDD) to different sub-groups in the sample.  The use of an RDD helps to account for potential endogeneity of the victory of a specific party's governor (see \cite{beland2015political}, \cite{lee2001electoral}).  The assumption used for identification is that in case of marginal victory, conditional on some observable factors, there are no other unobservables factors driving both the outcome of interest and the victory of a governor of a specific party.  

The descriptive analysis of earnings of black workers and white workers summarized in figures 1, 2, and 3 suggests that the earning distribution of White workers is bi-modal while  that of black workers is uni-model. The nature of the difference in the distribution makes the relevance of the average treatment effect as the parameter of interest questionable. Especially in a situation where the treatment effect has a very high likelihood of being heterogeneous. This implies that the estimation of the conditional expectation of difference estimators could be biased because of the non-symmetry of the treatment effects distribution. Indeed, the RDD creates a quasi-experiment around the cut-off of marginal victories, but the difference in the distribution of earnings could lead to an asymmetric distribution of treatment effects, which poses threats to significance testing as noted \cite{deaton2018understanding} for randomized control trials(RCT).\footnote{ \cite{deaton2018understanding} show that even in RCT, when the distribution of treatment effects are not symmetric the classical inference strategy may not be appropriate. RDD is usually considered as the quasi-experimental strategy closest to RTC, the over-representation of whites at the top of the distribution means that if the effect of governor depends on the level of income, the asymmetry is very likely.} 

To alleviate the consequence of potential biases in the estimates of the governors' effects using the full sample, we propose an estimation of the treatment effect on subgroups along earning and total hours-worked quantiles.\footnote{ The presence of masses probabilities on the distribution of earnings and total hours-worked prevents the used quantile regression for RDD as in \cite{fra12} which would have been our preferred estimation method.}  

For each quantile group, the following model is estimated

\begin{equation}\label{model}
    Y_{ist}= \beta_0 + \beta_{1,\tau} T_{st} +\beta_{2,\tau} T_{st} \times Black_{i} + G(MV_{st},Black_{i}, X_{ist} ) + \varepsilon_{ist}.
\end{equation}
  
The parameters of interest are $\beta_{2, \tau}$ with 
$\tau$ being a specific quantile. For an individual $i$, in the state $s$ at time $t$,  $Y_{ist}$ is the labour market outcome. $T_{st}$ is a dummy variable that takes the value 1 if a Democrat is in office as a governor in the state $s$ at time $t$.  The variable $Black_i$ indicates the ethnicity of the individual $i$, it takes the value 1 if the person is an African American and zero otherwise.  $MV_{st}$ is the victory margin in the last gubernatorial election in state $s$ prior to year $t$. $X_{ist}$ is the set of other observable characteristics that can affect the labor market outcome $Y_{ist}$. $G(.,.,.)$ represented different functional form specifications. The estimation of the parameters of Equation (\ref{model}) for each quantile group using an OLS estimator will deliver a consistent estimator under the assumption that the distribution of the labor market outcomes is similar in the subgroups. \\

\section{Discussion of Empirical Findings}

\subsection{Estimation Results}

Table \ref{tab:RDDH_E} to Table \ref{tab:RDDH_All} in Appendix~\ref{beau} show the results of the estimation of the causal effects of governors party affiliation decomposed in quantile sub-groups along two dimensions: earning and hours worked. Table  \ref{tab:RDDH_E} shows that when democrat governors’ policies are implemented, only black people in the high-earning quantile increase their average number of hours worked. However, blacks in the lower-earning quantile do not increase the average number of hours worked. The stagnation of the average number of hours worked in the lower-earning quantiles may be viewed as ``an hour trap’’, suggesting that incentives for blacks to work more hours are not boosted by Democrat policies in a complex economic and social environment which is shaped by the welfare system, the childcare, and the Medicare. There are several ways for understanding this finding. One possible interpretation is that all blacks in that earning quantile do not increase at all their hours worked when Democrat  policies are implemented. Another interpretation is that even if we assume some black in that low earning quantile increase the number of hours worked, this would be counterbalanced by another black in the same lower earning group that reduces the amount of time they used to work. The combined effect of these two forces can contribute to keeping the number of hours worked unchanged in the low earning quantiles. In that sense, the effects of the policies may not be effective in increasing the number of hours worked for black workers in the lower-earning quantiles. If a low-income black is already working too much to pay his/her bills, a government policy that increases childcare or paid sick leave may give incentive to that person to reduce his/her hours of works and enjoy more leisure time to have a more balanced life. In addition, the combination of income taxation combined with restrictions for people in the welfare system may also contribute to reduce work incentives and hold back hours worked for blacks.  There is literature that discusses how work fatigue accumulates over time as a consequence of work effort and can lead workers to reduce the number of hours worked \citep{don20,re18}. In the same way, health issues may create an incentive to work less, work fatigue may result in people who are already working several jobs or too many hours to reduce the number of hours worked.\footnote{Racial and ethnic minority populations have lower levels of access to medical care in the United States \citep{wiru00}. One recent shred of evidence on this issue is that Blacks are about half as likely as Whites to receive treatment for depression \citep{mcg2020}.}   

While it is commonly believed that democrats tend to put greater emphasis on addressing poverty and unemployment issues \citep{beland2015political,wo15,sie18,lei08,wri12},  our findings contrast and complement the study done by \cite{beland2015political} who using aggregate data found that under Democrat governors, blacks are more likely to work, participate in the labor market, and work more intensively. As \cite{beland2015political} mentioned, more work is needed in this area to understand the full extent of the role of political parties.

The findings in Table \ref{tab:RDDH_E} also show that the democrat governors’ policies only give incentive to black in the highest-earning quantile to increase their average hours worked.  State policies that subsidize childcare may have different effects on workers depending on the classification in terms of earning quantiles. It could be that childcare for instance allows black in the highest-earning quantile to take time away from childrearing to increase their hours worked.\footnote{The U.S. welfare system is composed of dozens of programs whose management is also shared by states and local governments, which actually deliver the services and contribute to funding. For instance, states can use funds related to the Temporary Assistance for Needy Families (TANF) to provide benefits to low-income families to help pay the costs of working, such as childcare or transportation.}

Table \ref{tab:RDDE_E} shows that none of the earning quantile taken individually experienced an increase in earnings after Democrat governors implemented their labor policies.  This may be because some workers who were already working too much decide to reduce their hours worked if for instance the minimum wage, childcare, or sick leave days become paid under democrat governors’ policies. In that case, the reduction in time may be driven by the need for a more balanced life. This may suggest the importance of empirically analyzing work fatigue as a factor to be taken into account in labor supply decisions along with democrat governors’ policies.\footnote{According to \cite{ricci07},  prevalence of fatigue is estimated at 37.9\% in the U.S. workforce.}

Table \ref{tab:RDDH_H} shows that only workers with the lowest number of hours worked to increase their number of hours worked after democrat governors implement their labor policies. This represents a total of 34\% of black workers who decide to increase their number of hours worked.

Table \ref{tab:RDDE_H} shows that in contrast to \cite{beland2015political} results, at the more micro subgroup level of hours worked, there has not been an increase in the earnings at all.  This may suggest that the increase in hours worked observed in the lowest quantile of hours worked in Table \ref{tab:RDDH_H} did not translate into a substantial increase in earnings. This could be viewed as a wage trap for that subgroup. It may be that the increase in hours worked is mainly driven by new workers. New workers who enter that subgroup after democrat governors policies are implemented may be new immigrants or people coming out of the welfare system or from prison whose marginal earnings are equal to or below the average wage of people who were already in that subgroup before the policy is implemented. This situation may be connected to the comparing marginal wage of new workers in a group with that of the average wage of worked already in the group before the implementation of the policy. If the marginal wage of new workers is less than the average wage of preexisting workers in that group, the average wage in that subgroup will not increase.\footnote{Even today, black workers are over-represented in low-wage entry-level jobs. In 2018, black workers made up 18\% of minimum wage workers despite being only 12.7\% of the population \citep{jec20}.}

Table \ref{tab:RDDE_H} shows that at the more micro levels in terms of subgroups related to the number of hours worked and earnings quantiles, there has not been an increase in the earnings at all.  This also could be explained by the effect of the marginal wage of new workers in a subgroup being less or equal than the average wage of preexisting workers in that subgroup, which leads to the average wage in that subgroup not increase after the democrat governor policies are implemented.

Table \ref{tab:RDDE_all} shows that there is no subgroup of black workers in terms of hours worked and earning quantiles that we're able to increase their earnings after the democrat governors’ policies are implemented.

Table \ref{tab:RDDH_All} brings additional light to Table \ref{tab:RDDH_H} by showing that workers in the subgroup composed of black workers in the first quantiles in hours worked and third quantiles in earnings ($Q_1$ Hours and $Q_3$ Earnings) and those in the second quantiles in terms of hours worked and third-fourth in earnings ($Q_2$ Hours and $Q_4$ Earnings) are the ones that increase their number of hours after the democrat governors’ policies are implemented. Again, the behavior of workers in $Q_1$ Hours and $Q_3$ Earnings shows that actually, only the workers who work the least and have earnings in the third quartile decide to increase their supply of labor. This raises the question of whether there is a ``catching-up with the Jones'' phenomena at play as those who work the least and are in the top highest-earning quantile do not increase their hours worked. The workers in the subgroup $Q_2$ Hours and $Q_4$ Earnings may be standard top earners who find it possible to increase the number of hours worked, given their flexible preferences between labor and leisure after the democrat governors' policies are implemented.

Since the findings demonstrate that subgroup heterogeneity with respect to  hours worked and earnings should be of interest to policymakers. The functioning of the American welfare system is a good one important element to have in mind when analyzing the full picture from our empirical results.

\subsection{Understanding why hours may increase or decrease while wage decrease or remain fixed for certain subgroups}

Let us start with explaining why wage may decrease or remain fixed. After the policy, if in a subgroup, there are new entrants (welfare leavers)  who decide to work full time but have an average wage that is lower than the previous average wage of that subgroup, then the average wage of that subgroup will decrease while the number of hours is increasing. Another explanation for the decrease in wage in a subgroup maybe that new entrants (welfare leavers) do not work for four quarters in a row, signaling a potential problem with employee retention and stability. Now let us provide a possible explanation for a decrease in hours as it related to the welfare system. It may be after that after the state policy, the number of people previously in the subgroup but that are not able to keep stable jobs is more important quantitatively than people (welfare leavers) entering that subgroup that are working full time.\footnote{On the period 1981-2017, \cite{wri19} found that black/white displacement disparities have grown over time, with excess Black displacement doubling for women and tripling for men since the 1990s. They found that during the 1990s, being Black replaced lacking a college degree as the better predictor of displacement.}

\subsection{Earning trap and hours worked trap}

The results have shown that the average wage of several groups does not change when the policies are implemented. This suggests several subgroups experience a ``wage trap'' , especially the group with the lower average wage. The “low-wage trap” shown in the estimations may be understood as a situation where those already in low-paid work are not able to increase their income when democrat governors implement their policies. The combination of income taxes, social security contributions, and benefits withdrawal may “tax away” all or a large part of any wage gain.\footnote{The growing stratification economics literature contends that social prejudices influence productivity in labor markets \citep{da17}. Historically, black workers are less likely to work in jobs that provide retirement benefits. As far as back during the interwar 1919-1943, \cite{war88} found that black workers had less access to stable jobs (high-turnover jobs) and complained about several kinds of discrimination in job assignments. }

The estimations have also shown that the number of hours worked does not change after the democrat governors implement their policies. This suggests the existence of a ``hours worked trap’’  referring to a situation where the number of hours worked does not change for a certain subgroup of workers.  Income restrictions related to the welfare system may be an important factor to be considered in explaining the hours' trap observed for certain subgroups. Some restrictions in the welfare system can entertain situations where increasing the number of hours worked is judged by certain subgroups to be not to not optimal from a well-being standpoint due to expected income-related benefits which would be lost upon working a full-time paid job. 

\subsection{State policies are needed to improve the income gains of those who have left welfare for work}

The benefits and taxes can create a wage floor below which increasing the number of hours worked for certain subgroups does not bring any financial gain in the short term.\footnote{\cite{dere20} provide a piece of causal evidence on how minimum wage policy can play a critical role in reducing the income disadvantage suffered by black workers in the United States.} Taxes and benefits while affecting earnings shape work incentives.
If earnings gains are insufficient to counter reductions in benefits, the number of hours worked may decrease or be stagnant for certain subgroups. Therefore, there is a need for a comprehensive strategy to assist individuals in the subgroups where both earnings and hours worked do not change after the democrat governors’ policies are implemented.  In assisting these subgroups, developing new innovative ways to deal with issues related to low levels of education and job skills,\footnote{As shown by \cite{car17}, public sector discrimination in denying black public schools the same quality available to white public schools played a predominant in determining wage differences for black and white workers during the Jim-crow era in the United States.} weak public transportation systems, health problems, substance abuse, and domestic violence, and discrimination in labor markets should be at the forefront of a strategy that pursues poverty reduction while maximizing social well-being for black workers.\footnote{\cite{dari18} highlight complexities surrounding commonly held myths about economic issues faced by black households in the United States.}

\section{Conclusion}
Understanding how  black workers in the United States  are affected by US state policies in labor markets can shed light on new ways to improve or implement public policymaking. While there is substantial literature on the aggregate impact of labor policies on black workers, little is known on the distributional impact of state policies among black workers in the labor market. In this paper, we have contributed to this debate by highlighting the distributional impact of Democrat governors' policies across twelve subgroups of black workers sorted on two dimensions: earnings quantiles and hours worked. With respect to wages, we found that none of these subgroups experienced wage increase after Democrat governor’s policies are implemented. With respect to hours worked, only two subgroups increase their hours worked. These are workers in the subgroup composed of Black workers in the first quantile in hours worked and the third quantile in earnings ($Q_1$ Hours and  $Q_3$ Earnings) and  the ones in the subgroup composed  of black workers in  the second quantile in terms of hours worked and third-fourth in terms of  earnings ($Q_2$ Hours and  $Q_4$ Earnings). In contrast, there is a subgroup that decreases their average number of hours worked. This subgroup is composed of people who are both in the lowest-earning quantile and highest hours worked quantile. The causal empirical estimations highlight that the vast majority of subgroups of black workers are not affected at all by Democrat governors’ policies, suggesting the possible existence of structural factors in the labor markets that contribute to create and keep a wage trap and/or hour worked trap for most of the subgroups of black workers. Our results also suggest the importance of incorporating a wide range of heterogeneity at the household level while designing economic theoretical models aiming at understanding the impact of state policies on black workers in the US. 

There is currently a great deal of discussion on how to improve the labor markets outcomes for minorities in the US. One important insight from our paper is that the distribution of earnings and hours worked among black workers can have considerable significance on how the Democrat governor’s labor policies affect them.  Our results suggest that even within minority ethnic groups in the US, accounting for heterogeneity in the distribution impacts of policies implemented should not be overlooked and can provide more knowledge on how to improve the social well-being overall.  In designing policies at the state level, it is important to take into account some characteristics of labor markets suggesting that jobs with more black workers do tend to pay less than other jobs \citep{huf04}, and according to \cite{bos90} the hypothesis that workers in the primary sector engage in less frequent job changes than workers in the secondary sector is supported for white workers but not for black workers, which negatively affect their work experience. Black Americans have substantially lower rates of upward mobility and higher rates of downward mobility than whites \citep{che20}.\footnote{Historically, the origins of economic issues faced by black households could be traced back to the failure to provide the formerly enslaved with the promised 40-acre land grants in the immediate aftermath of the Civil War \citep{wi21}. The dynamic of labor markets is not immune to the historical context in which existing social, economic, and economic forces have been shaped. One example is provided by  \cite{ra90} who questions the functioning of the labor markets while accounting for the spillover effects of past racial institutions.} Along the same lines, only 10\% of black households could be viewed as ``solidly’’ middle-class \citep{wi21} in the United States. Another unusual disadvantage faced by black workers is related to the nexus between the penal system and the labor market \citep{wes18}.  One in three Black men will be incarcerated over his lifetime \citep{jec20}, and this will disproportionally hinder their future earning prospects \citep{wes05}.\footnote{The consequences of mass incarceration on black communities are far reaching - including greater health disparities, the destruction of the black family, greater obstacles to human capital investment, and food insecurity \citep{cox18,cox16}.} Finally, the staggering job losses sparked by the coronavirus pandemic have disproportionately affected Black communities and there is a concern that the  COVID-19 recession is expected to be felt by  Black workers even once it concludes \citep{hot21}.

\newpage

\bibliographystyle{ecta}
\bibliography{ref1}
\newpage
\appendix

\section{Appendix: Figures 5 to 11 \label{Hallelujah}}

\begin{figure}[H]
    \centering
     \caption{Distribution of income by race Under a Democrat}
    \includegraphics[scale=0.8]{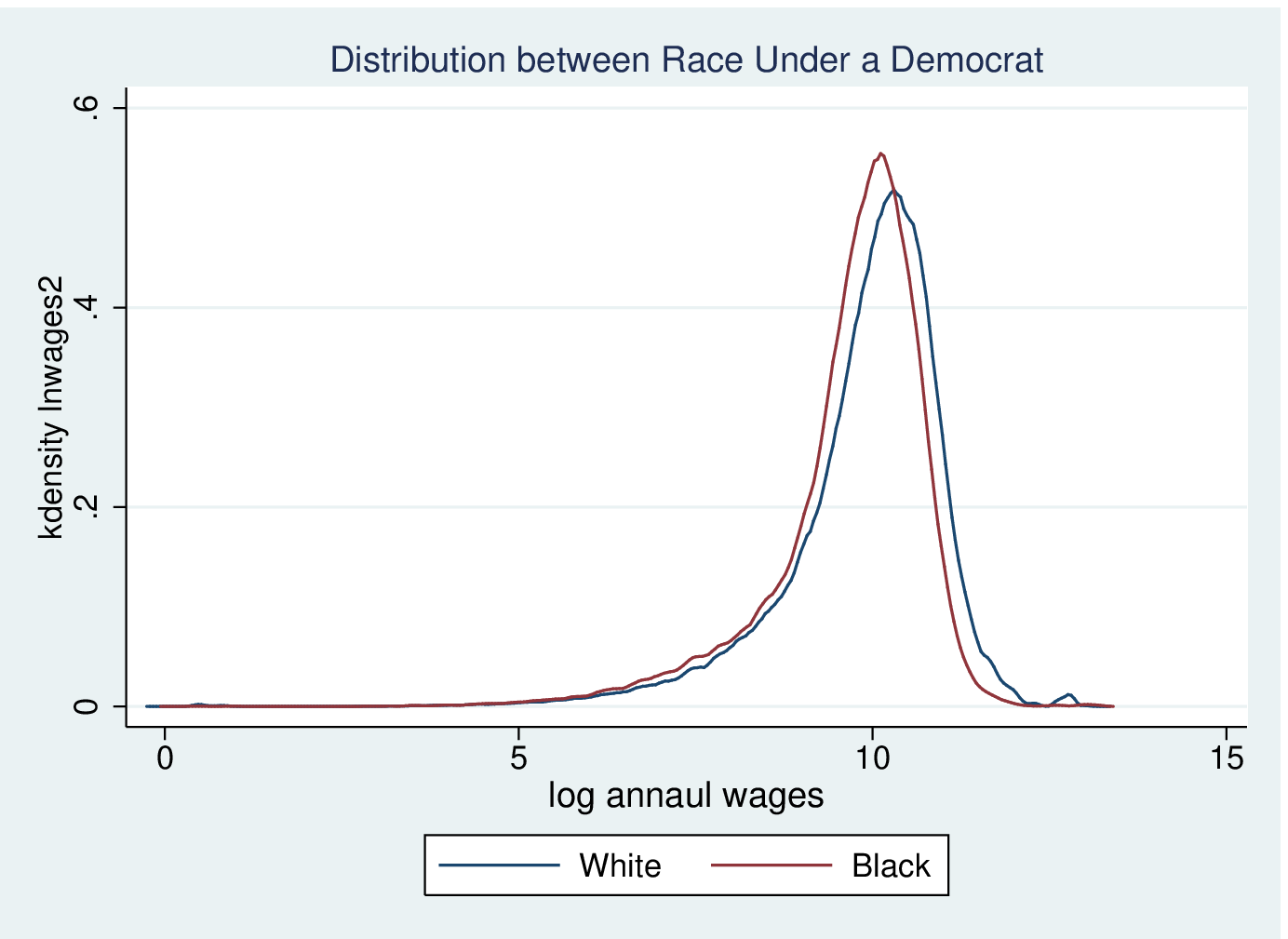}
     \label{fig: race_onlyd}
\end{figure}

\begin{figure}[H]
    \centering
     \caption{Distribution of income by race Under a Republican}
    \includegraphics[scale=0.8]{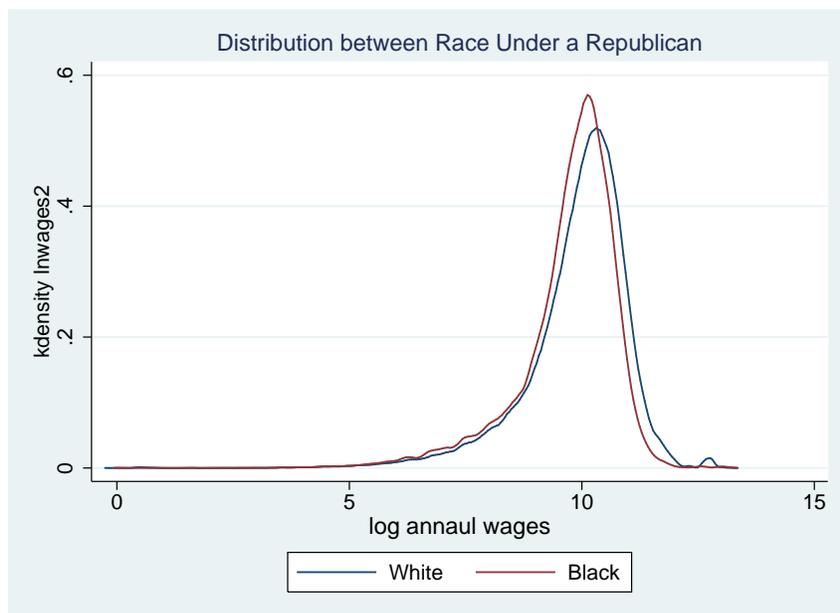}
     \label{fig: race1_92}
\end{figure}

\begin{figure}[H]
    \centering
     \caption{Distribution of income of whites over time under Dems vs Reps.}
    \includegraphics[scale=0.8]{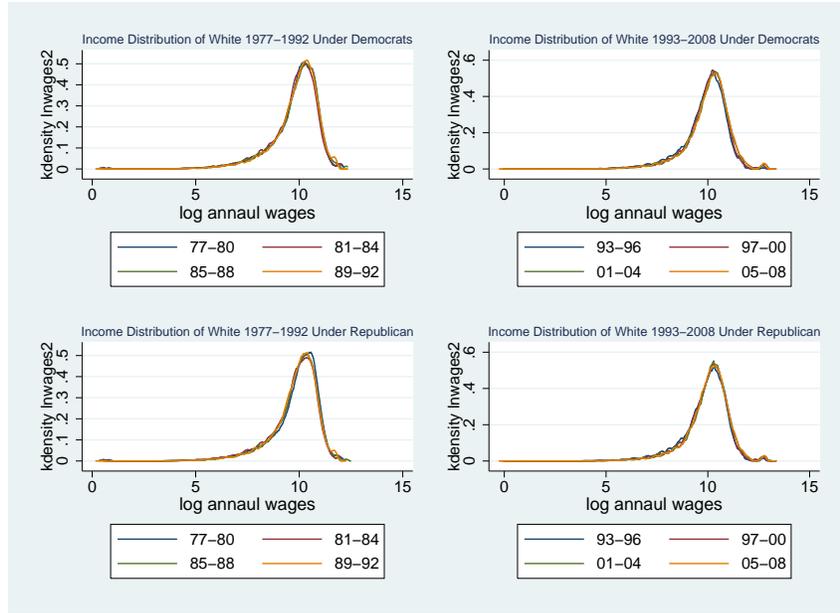}
     \label{fig: race1_92}
\end{figure}

\begin{figure}[H]
    \centering
     \caption{Distribution of income of blacks over time under Dems vs Reps.}
    \includegraphics[scale=0.8]{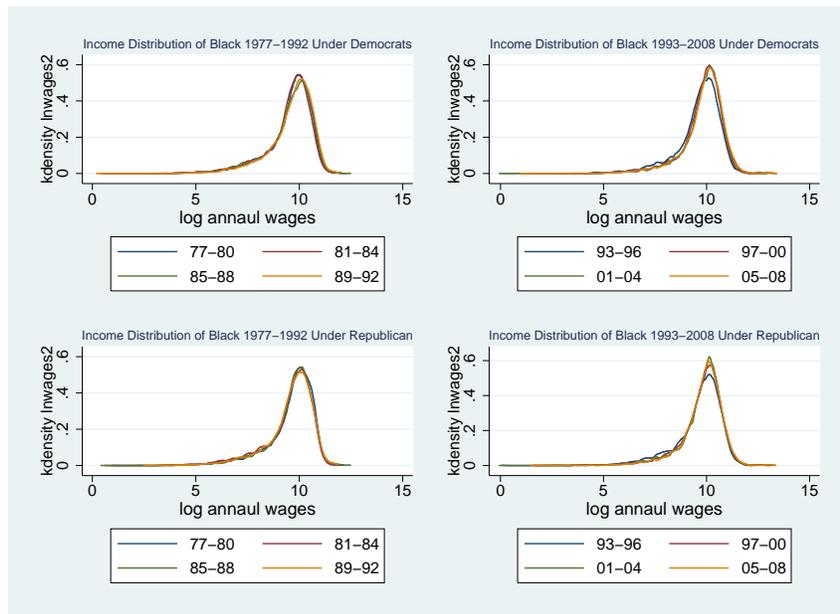}
     \label{fig: race1_92}
\end{figure}

\begin{figure}[H]
    \centering
     \caption{Distribution of income by race 4-year split 1993--2008 (guess Race=1 is White)}
    \includegraphics[scale=0.9]{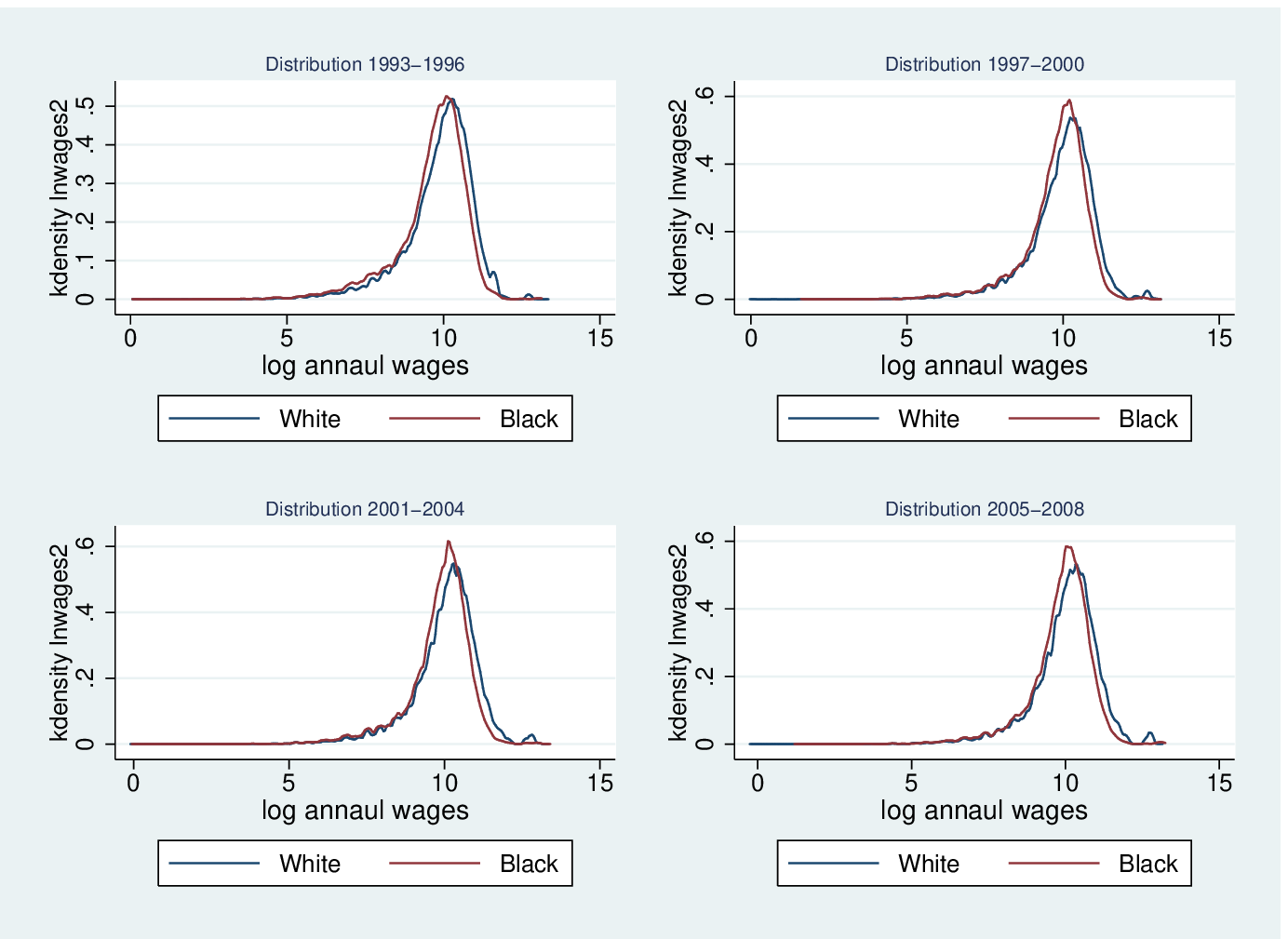}
     \label{fig: race1_08}
\end{figure}

\begin{figure}[htpb!]
    \centering
    \caption{Margin of Democratic Victory and the proportion of black (left on each panel) and white (right on each panel) who worked.}
    \includegraphics{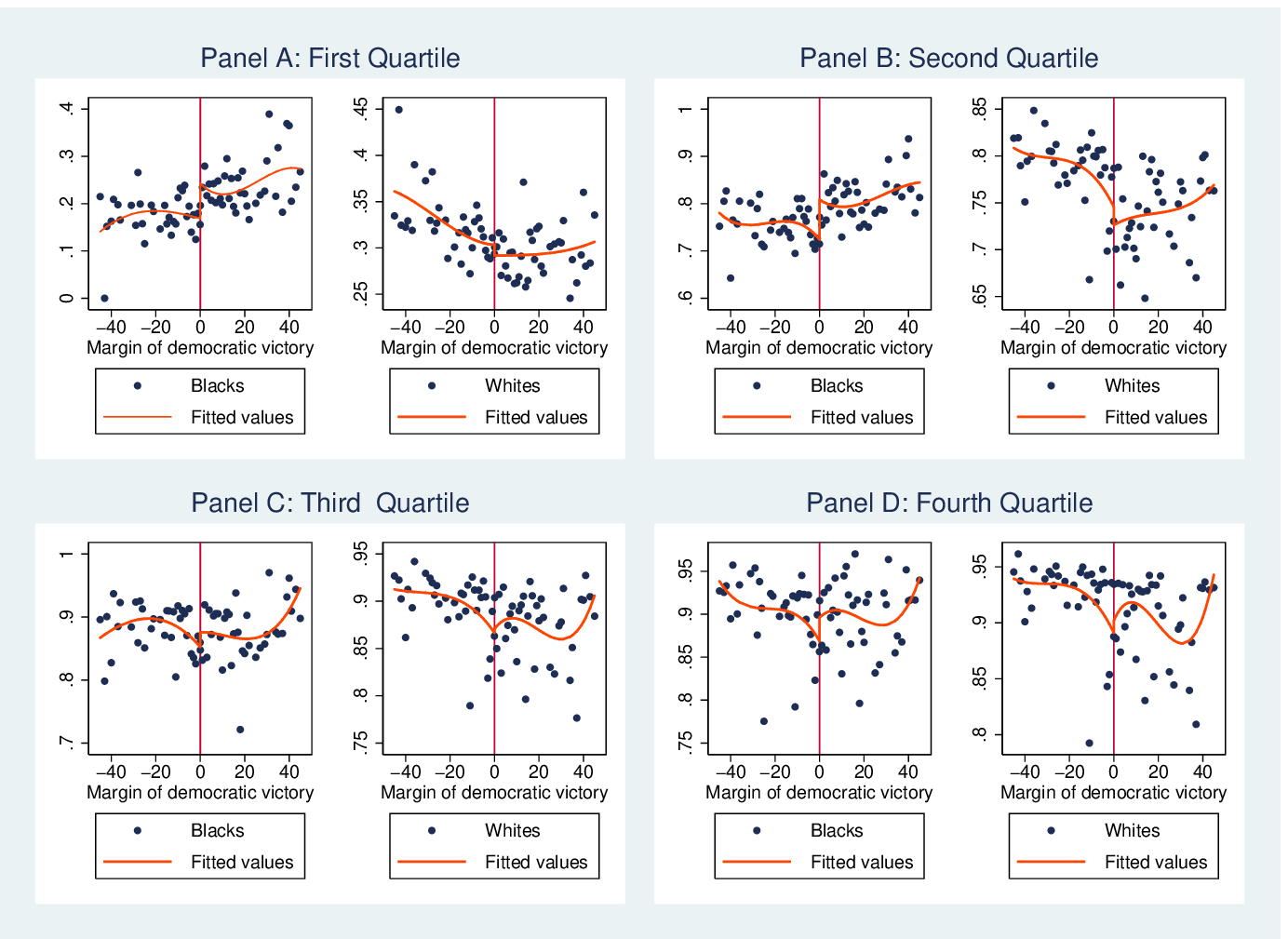}
    
    \label{fig:rdgraph_worked}
\end{figure}

\begin{figure}[htpb!]

\caption{Margin of Democratic Victory and log wages for black (left on each panel) and white (right on each panel). The analysis is by Earning quartile.}
    \centering
    \includegraphics{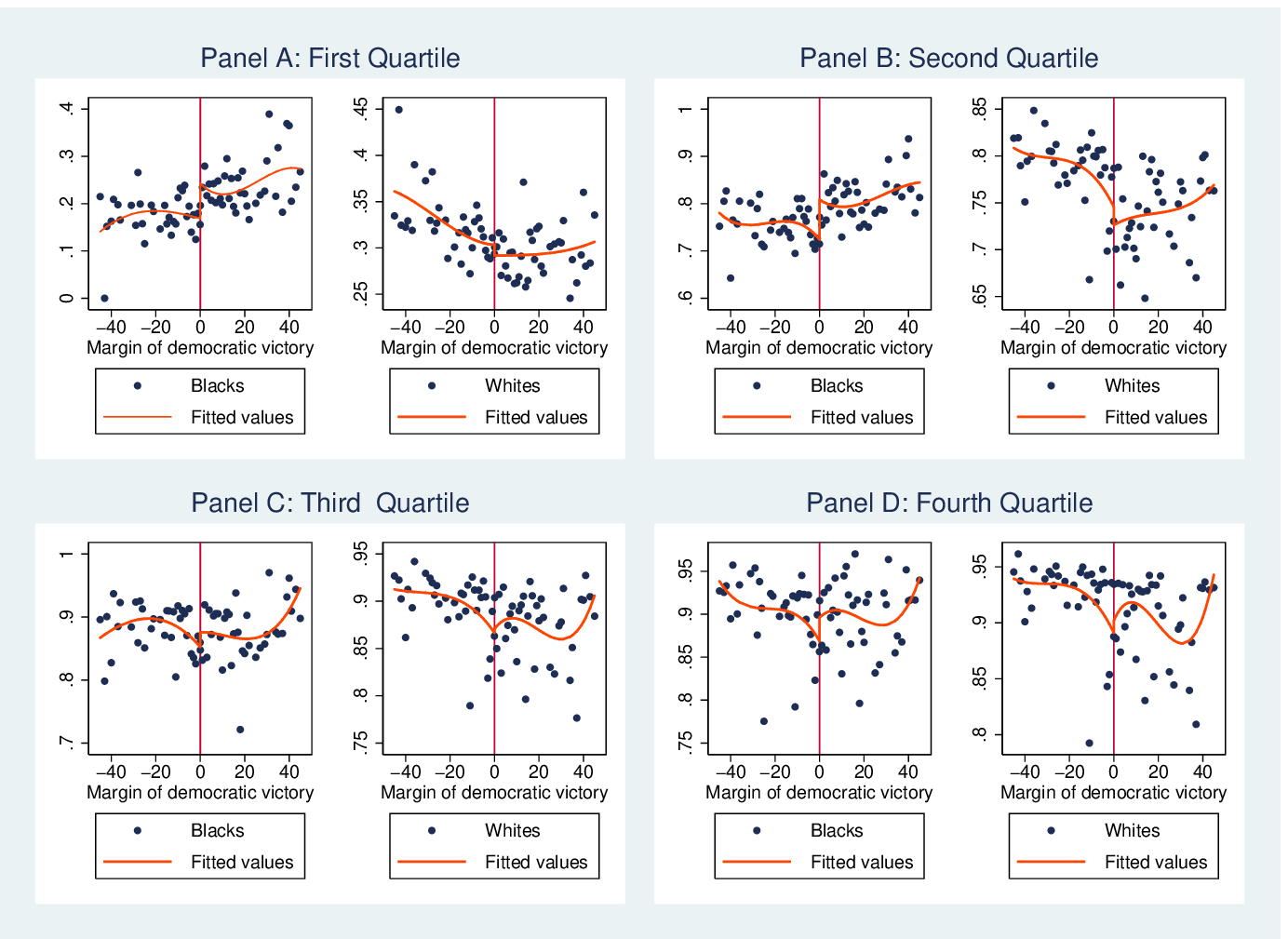}
    
    \label{fig:rdgraph_wage}
\end{figure}
\FloatBarrier

\section{Appendix: Tables 3 to 10 \label{beau}}
\vspace{.5cm}

\FloatBarrier
\begin{table}[H]
    \centering
        \caption{Two-sample Kolmogorov-Smirnov test for equality of distribution functions}
    \begin{tabular}{l@{\hskip 0.4in} c@{\hskip 0.3in} c}\toprule
   Smaller group & D & P-value \\ \hline
    && \\ [-0.4em]
     Blacks      &  0.0005 &   0.904 \\ [0.3em]
     White       & -0.1127 &   0.000\\ [0.3em]
     Combined    &  0.1127 &   0.000 \\
     \bottomrule
    \end{tabular}
    \label{tab:my_label}
\end{table}
\vspace{0.5cm}

\begin{table}[btp]    \footnotesize  
\caption{Proportion of blacks by quartile of earning / hours worked}
    \label{tab:my_label}
    \centering
    \begin{threeparttable}
{
\def\sym#1{\ifmmode^{#1}\else\(^{#1}\)\fi}
\begin{tabular}{l@{\hskip 0.4in}*{5}{c}}
\hline\hline
          &              \multicolumn{4}{c}{Quartile (Total Earnings)} \\ \cmidrule(rl){2-5}
 Quartile (Hours)              &\multicolumn{1}{c}{$Q_1$ Earning }&\multicolumn{1}{c}{$Q_2$ Earning}&\multicolumn{1}{c}{ $Q_3$ Earning}&\multicolumn{1}{c}{$Q_4$ Earning}&\\
\hline
&&& \\ [-0.3em]

$Q_1$ Hours &       12.22\%         &      9.83\%         &   8.62\% &     7.36\%         \\
            &      (5.63\%)         &     (13.92\%)         &      (1.68\%)         &     (0.47\%)         \\[1em]

 $Q_2$ Hours &      11.21\%         &      13.16\%         &       11.74\% &       8.35\% \\
            &     (0.38\%)         &     (21.72\%)         &      (28.35\%)         &      (15.43\%)         \\[1em]

$Q_{3,4}$ Hours  &      2.72\% &    8.30\%         &       6.97\%    &     4.30\%         \\
            &     (0.04\%)         &     (1.79\%)         &      (4.72\%)         &      (5.87\%)         \\
\hline\hline
\end{tabular}}
\begin{tablenotes} \footnotesize
\item {\em Notes:} Percentages in parenthesis are computed as the number of blacks in a given quartile combination divided by the total number black in the entire sample. In contrast, percent without parenthesis is the with group proportion of blacks.
\end{tablenotes}
\end{threeparttable}
\end{table}
\vspace{0.5cm}

\begin{table}[htbp!]
\centering   \caption{RD estimates for total hours worked: Earnings as Quartile  }
\label{tab:RDDH_E}
    \begin{threeparttable} 
{
\def\sym#1{\ifmmode^{#1}\else\(^{#1}\)\fi}
 \footnotesize  \begin{tabular}{l*{5}{c}}
\hline\hline
            &\multicolumn{1}{c}{(1)}&\multicolumn{1}{c}{(2)}&\multicolumn{1}{c}{(3)}&\multicolumn{1}{c}{(4)}&\multicolumn{1}{c}{(5)}\\
            &\multicolumn{1}{c}{Beland (2015, Table 2)}&\multicolumn{1}{c}{$Q_1$ Earnings}&\multicolumn{1}{c}{$Q_2$ Earnings}&\multicolumn{1}{c}{$Q_3$ Earnings}&\multicolumn{1}{c}{$Q_4$ Earnings}\\
\hline
&&& \\ [-0.3em]
Democrat      &      -0.006         &      -0.025         &       0.004         &       0.005         &       0.002         \\
            &     (-1.06)         &     (-0.74)         &      (0.53)         &      (1.38)         &      (1.05)         \\
[1em]
Black $\times$ Democrat &       0.038\sym{**} &       0.035         &      -0.011         &       0.005         &       0.013\sym{**} \\
            &      (2.09)         &      (0.35)         &     (-0.54)         &      (0.58)         &      (2.32)         \\
[1em]
Black      &      -0.065\sym{***}&      -0.088         &       0.028\sym{**} &      -0.009\sym{**} &      -0.041\sym{***}\\
            &     (-5.61)         &     (-1.23)         &      (2.39)         &     (-2.59)         &    (-12.56)         \\
[1em]
Constant      &      -4.244\sym{***}&       3.420\sym{***}&       0.292         &       5.852\sym{***}&       5.782\sym{***}\\
            &    (-10.49)         &      (3.40)         &      (0.79)         &     (43.95)         &     (29.14)         \\
\hline
\(N\)       &     1912653         &       94770         &      606923         &      605602         &      605358         \\ [0.3em]
\hline\hline
\end{tabular}
}
\begin{tablenotes} \footnotesize
\item {\em Notes:} Quartiles are governed by wages. Control variables include highest level of education, marital status, age, age two, age three, age four, a female dummy, state fixed effects, and year fixed effects. Outcome variables are expressed in log form. Results are clustered at the state level. {\textit{t} statistics in parentheses}.
\item *** Significant at the 1 percent level.
\item \hspace{0.1cm} ** Significant at the 5 percent level.
\item \hspace{0.3cm} * Significant at the 10 percent level.
\end{tablenotes}
\end{threeparttable}
    \label{tab:my_label}
\end{table}
\vspace{0.5cm}

\begin{table}[htpb!]     \caption{RD Estimate Effect for Earnings: Earnings as Quartile  }
    \label{tab:RDDE_E}
\footnotesize     \centering
\begin{threeparttable}
{
\def\sym#1{\ifmmode^{#1}\else\(^{#1}\)\fi}
\begin{tabular}{l*{5}{c}}
\hline\hline
            &\multicolumn{1}{c}{(1)}&\multicolumn{1}{c}{(2)}&\multicolumn{1}{c}{(3)}&\multicolumn{1}{c}{(4)}&\multicolumn{1}{c}{(5)}\\
            &\multicolumn{1}{c}{Beland (2015, Table A2)}&\multicolumn{1}{c}{$Q_1$ Earning}&\multicolumn{1}{c}{$Q_2$ Earning}&\multicolumn{1}{c}{$Q_3$ Earning}&\multicolumn{1}{c}{$Q_4$ Earning}\\
\hline
&&& \\ [-0.3em]
Democrat      &      -0.024\sym{**} &      -0.038         &       0.004         &      -0.002         &      -0.005         \\
            &     (-2.16)         &     (-1.60)         &      (0.78)         &     (-0.82)         &     (-0.92)         \\
[1em]
Black $\times$ Democrat &       0.050\sym{*}  &      -0.011         &      -0.001         &       0.004         &       0.009         \\
            &      (1.99)         &     (-0.18)         &     (-0.04)         &      (0.66)         &      (0.68)         \\
[1em]
Black       &      -0.164\sym{***}&       0.022         &      -0.007         &      -0.016\sym{***}&      -0.073\sym{***}\\
            &    (-10.34)         &      (0.59)         &     (-0.66)         &     (-3.15)         &     (-9.24)         \\
\hline
\(N\)       &     1921186         &       94873         &      608775         &      609851         &      607687         \\
\hline\hline
\end{tabular}
}
\begin{tablenotes} \footnotesize
\item {\em Notes:} Quartiles are governed by wages. Control variables include highest level of education, marital status, age, age two, age three, age four, a female dummy, state fixed effects, and year fixed effects. Outcome variables are expressed in log form. Results are clustered at the state level. {\textit{t} statistics in parentheses}.
\item *** Significant at the 1 percent level.
\item \hspace{0.1cm} ** Significant at the 5 percent level.
\item \hspace{0.3cm} * Significant at the 10 percent level.
\end{tablenotes}
\end{threeparttable}
\end{table}
\vspace{0.5cm}

\begin{table}[htpb!]    \footnotesize  \caption{RD estimates for total hours worked: Hours as Quartile }
    \label{tab:RDDH_H}
    \centering
    \begin{threeparttable}
{
\def\sym#1{\ifmmode^{#1}\else\(^{#1}\)\fi}
\begin{tabular}{l*{5}{c}}
\hline\hline
            &\multicolumn{1}{c}{(1)}&\multicolumn{1}{c}{(2)}&\multicolumn{1}{c}{(3)}&\multicolumn{1}{c}{(4)}&\\
            &\multicolumn{1}{c}{Beland (2015, Table A2)}&\multicolumn{1}{c}{$Q_1$ Hours}&\multicolumn{1}{c}{$Q_2$ Hours}&\multicolumn{1}{c}{$Q_3$ Hours}& \\
\hline
&&& \\ [-0.3em]

Democrat      &      -0.006         &      -0.006         &      -0.000         &      -0.000         \\
            &     (-1.06)         &     (-0.52)         &     (-0.29)         &     (-0.32)         \\
[1em]
Black $\times$ Democrat &       0.038\sym{**} &       0.080\sym{**} &       0.001         &       0.007         \\
            &      (2.09)         &      (2.33)         &      (0.32)         &      (0.79)         \\
[1em]
Black      &      -0.065\sym{***}&      -0.106\sym{***}&       0.006\sym{***}&       0.007         \\
            &     (-5.61)         &     (-4.23)         &      (4.28)         &      (1.11)         \\
[1em]
            &    (-10.49)         &      (3.59)         &    (213.90)         &     (93.24)         \\
\hline
\(N\)       &     1912653         &      387536         &     1103159         &      421958         \\
\hline\hline
\end{tabular}
}
\begin{tablenotes} \footnotesize
\item {\em Notes:} Quartiles are governed by total hours worked. Control variables include highest level of education, marital status, age, age two, age three, age four, a female dummy, state fixed effects, and year fixed effects. Outcome variables are expressed in log form. Results are clustered at the state level. {\textit{t} statistics in parentheses}.
\item *** Significant at the 1 percent level.
\item \hspace{0.1cm} ** Significant at the 5 percent level.
\item \hspace{0.3cm} * Significant at the 10 percent level.
\end{tablenotes}
\end{threeparttable}
\end{table}
\vspace{0.5cm}

\begin{table}[htpb!]    \footnotesize  \caption{RD Estimate Effect for Earnings:  Hours as Quartile }
    \label{tab:RDDE_H}
    \centering
    \begin{threeparttable}
{
\def\sym#1{\ifmmode^{#1}\else\(^{#1}\)\fi}
\begin{tabular}{l*{5}{c}}
\hline\hline
            &\multicolumn{1}{c}{(1)}&\multicolumn{1}{c}{(2)}&\multicolumn{1}{c}{(3)}&\multicolumn{1}{c}{(4)}&\\
            &\multicolumn{1}{c}{Beland (2015, Table A2)}&\multicolumn{1}{c}{$Q_1$ Hours}&\multicolumn{1}{c}{$Q_2$ Hours}&\multicolumn{1}{c}{$Q_3$ Hours}&\\
\hline
&&& \\ [-0.3em]

Democrat      &      -0.024\sym{**} &      -0.033\sym{*}  &      -0.023\sym{***}&      -0.010         \\
            &     (-2.16)         &     (-1.88)         &     (-2.86)         &     (-0.79)         \\
[1em]
Black $\times$ Democrat &       0.050\sym{*}  &       0.063         &       0.033         &      -0.011         \\
            &      (1.99)         &      (1.29)         &      (1.60)         &     (-0.36)         \\
[1em]
Black      &      -0.164\sym{***}&      -0.160\sym{***}&      -0.100\sym{***}&      -0.120\sym{***}\\
            &    (-10.34)         &     (-4.28)         &     (-6.51)         &     (-6.07)         \\
\hline
\(N\)       &     1921186         &      396069         &     1103159         &      421958         \\
\hline\hline
\end{tabular}
}
\begin{tablenotes} \footnotesize
\item {\em Notes:} Quartiles are governed by total hours worked. Control variables include highest level of education, marital status, age, age two, age three, age four, a female dummy, state fixed effects, and year fixed effects. Outcome variables are expressed in log form. Results are clustered at the state level. {\textit{t} statistics in parentheses}.
\item *** Significant at the 1 percent level.
\item \hspace{0.1cm} ** Significant at the 5 percent level.
\item \hspace{0.3cm} * Significant at the 10 percent level.
\end{tablenotes}
\end{threeparttable}
\end{table}


\begin{table}[H]    \footnotesize  \caption{Dependent variable, Wages (Earnings)  }
    \label{tab:RDDE_all}
    \centering
    \begin{threeparttable}
{
\def\sym#1{\ifmmode^{#1}\else\(^{#1}\)\fi}
\begin{tabular}{l@{\hskip 0.4in}*{5}{c}}
\hline\hline
          &              \multicolumn{4}{c}{Quartile (Total Earnings)} \\ \cmidrule(rl){2-5}
 Quartile (Hours)               &\multicolumn{1}{c}{$Q_1$ Earning}&\multicolumn{1}{c}{$Q_2$ Earnings}&\multicolumn{1}{c}{$Q_3$ Earnings}&\multicolumn{1}{c}{$Q_4$ Earnings}&\\
\hline
&&& \\ [-0.3em]

 $Q_1$ Hours &       0.003         &       0.004         &       0.020         &      -0.130         \\ 
            &      (0.06)         &      (0.19)         &      (0.93)         &     (-1.30)         \\ [1em]
 $Q_2$ Hours  &      -0.348         &       0.004         &       0.005         &       0.013         \\ 
            &     (-1.35)         &      (0.38)         &      (0.81)         &      (0.97)         \\ [1em]

 $Q_3$ Hours &       0.591         &      -0.048         &      -0.001         &       0.011         \\ 
            &      (0.34)         &     (-0.92)         &     (-0.07)         &      (0.51)         \\
\hline\hline
\end{tabular}
}
\begin{tablenotes} \footnotesize
\item {\em Notes:} Control variables include highest level of education, marital status, age, age two, age three, age four, a female dummy, state fixed effects, and year fixed effects. Outcome variables are expressed in log form. Results are clustered at the state level. {\textit{t} statistics in parentheses}.
\item *** Significant at the 1 percent level.
\item \hspace{0.1cm} ** Significant at the 5 percent level.
\item \hspace{0.3cm} * Significant at the 10 percent level.
\end{tablenotes}
\end{threeparttable}
\end{table}
\vspace{0.5cm}

\begin{table}[H]    \footnotesize  \caption{Dependent variable, Total Hours}
    \label{tab:RDDH_All}
    \centering
    \begin{threeparttable}
{
\def\sym#1{\ifmmode^{#1}\else\(^{#1}\)\fi}
\begin{tabular}{l@{\hskip 0.4in}*{5}{c}}
\hline\hline
          &              \multicolumn{4}{c}{Quartile (Total Earnings)} \\ \cmidrule(rl){2-5}
 Quartile (Hours)              &\multicolumn{1}{c}{$Q_1$ Earning}&\multicolumn{1}{c}{$Q_2$ Earnings}&\multicolumn{1}{c}{$Q_3$ Earnings}&\multicolumn{1}{c}{$Q_4$ Earnings}&\\
\hline
&&& \\ [-0.3em]

 $Q_1$ Hours &       0.069         &      -0.003         &       0.129\sym{***}&      -0.010         \\
            &      (0.83)         &     (-0.16)         &      (2.81)         &     (-0.06)         \\[1em]

$Q_2$ Hours  &      -0.035         &      -0.008         &       0.001         &       0.004\sym{**} \\
            &     (-0.97)         &     (-1.49)         &      (0.21)         &      (2.08)         \\[1em]

$Q_3$ Hours  &      -0.293\sym{**} &      -0.002         &       0.008         &       0.014         \\
            &     (-2.45)         &     (-0.11)         &      (0.77)         &      (1.15)         \\
\hline\hline
\end{tabular}}

\begin{tablenotes} \footnotesize
\item {\em Notes:} Control variables include highest level of education, marital status, age, age two, age three, age four, a female dummy, state fixed effects, and year fixed effects. Outcome variables are expressed in log form. Results are clustered at the state level. {\textit{t} statistics in parentheses}.
\item *** Significant at the 1 percent level.
\item \hspace{0.1cm} ** Significant at the 5 percent level.
\item \hspace{0.3cm} * Significant at the 10 percent level.
\end{tablenotes}
\end{threeparttable}
\end{table}
\end{document}